\documentclass[aps, pra, twocolumn, tightenlines, letterpaper, amsmath, amssymb, preprintnumbers, floatfix, longbibliography, nofootinbib]{revtex4-2}

\usepackage{dsfont}
\usepackage{amsmath}
\usepackage{amssymb}
\usepackage{physics}
\usepackage{tabu}
\usepackage{tabularx}
\usepackage{bm}
\usepackage{tikz}
\usetikzlibrary{quantikz}
\usepackage{graphicx}
\usepackage{placeins}
\usepackage{textgreek}
\usepackage{multirow}
\usepackage{makecell}
\usepackage{soul}
\usepackage{diagbox}
\usepackage{xcolor}
\usepackage[normalem]{ulem}

\makeatletter
\newcommand{\fmarki}{*}
\newcommand{\fmarkii}{\ensuremath{\dagger}}
\newcommand{\fmarkiii}{\ensuremath{\ddagger}}
\newcommand{\fmarkiv}{\ensuremath{\mathsection}}
\newcommand{\fmarkv}{\ensuremath{\mathparagraph}}
\newcommand{\fmarkvi}{\ensuremath{\|}}

\def\si{{}^1\kern-.14em S_0}
\def\siii{{}^3\kern-.14em S_1}
\def\piii{{}^3\kern-.14em P_1}
\def\diii{{}^3\kern-.14em D_1}

\newcommand{\Moller}{M{\o}ller }

\def\@fnsymbol#1{{\ifcase#1\or \fmarki\or \fmarkii\or \fmarkiii\or \fmarkiv\or \fmarkv\or \fmarkvi \else\@ctrerr\fi}}
\makeatother

\renewcommand{\fmarkvi}{\$}

\usepackage[hypertexnames=false]{hyperref}
\hypersetup{
    colorlinks=true,       
    linkcolor=blue,          
    citecolor=blue,        
    filecolor=blue,      
    urlcolor=blue           
}

\newcolumntype{Y}{>{\centering\arraybackslash}X}

\makeatletter
\pretocmd\frontmatter@thefootnote{\color{black}}{}{}
\makeatother

\usepackage{orcidlink}

\begin{document}

 \begin{figure}
   \vskip -1.cm
   \leftline{\includegraphics[width=0.15\textwidth]{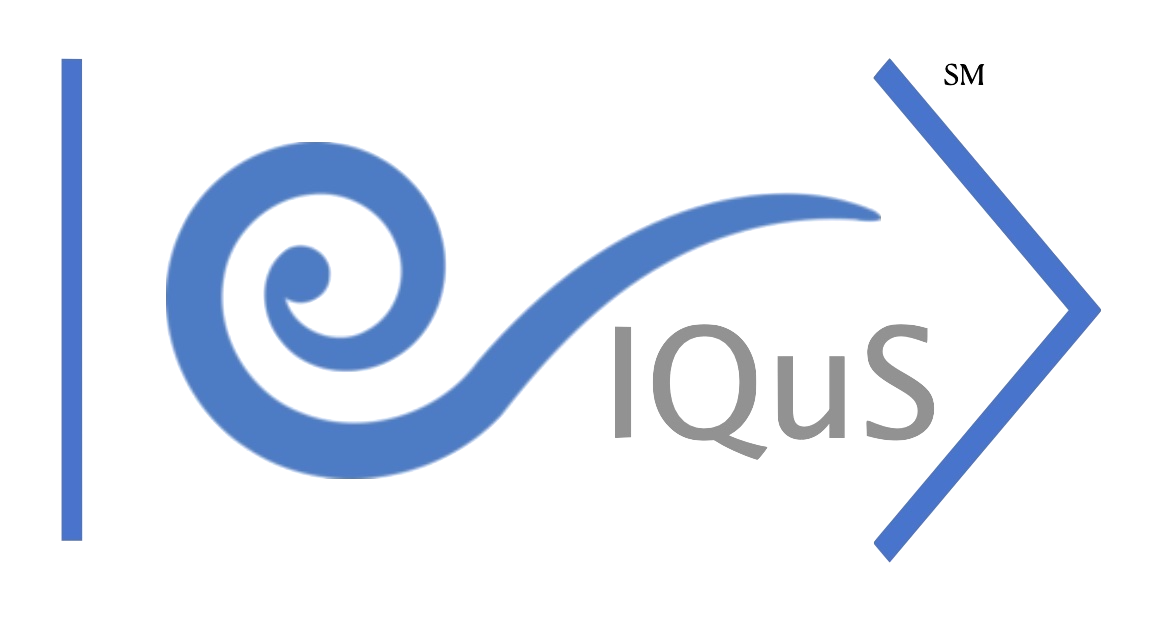}}
 \end{figure}

\title{Anti-Flatness and Non-Local Non-Stabilizerness in Two-Particle Scattering Processes}

\author{Caroline E.~P.~Robin\,\orcidlink{0000-0001-5487-270X}}
\email{crobin@physik.uni-bielefeld.de}
\affiliation{Fakult\"at f\"ur Physik, Universit\"at Bielefeld, D-33615, Bielefeld, Germany}
\affiliation{GSI Helmholtzzentrum f\"ur Schwerionenforschung, Planckstra{\ss}e 1, 64291 Darmstadt, Germany}

\author{Martin J.~Savage\,\orcidlink{0000-0001-6502-7106}}
\email{mjs5@uw.edu}
\thanks{On leave from the Institute for Nuclear Theory.}
\affiliation{InQubator for Quantum Simulation (IQuS), Department of Physics, University of Washington, Seattle, WA 98195, USA.}

\preprint{IQuS@UW-21-111}
\date{\today}

\begin{abstract}
\noindent
Non-local non-stabilizerness and anti-flatness provide a measure of the quantum complexity
in the wavefunction of a physical system.
Supported by entanglement, they cannot be removed by local unitary operations, 
thus providing basis-independent measures,
and sufficiently large values underpin the need for quantum computers 
in order to perform precise simulations of the system 
at scale.
Towards a better understanding of the quantum-complexity generation by fundamental interactions, the building blocks of many-body systems, 
we consider non-local non-stabilizerness and anti-flatness in 
two-particle scattering processes, 
specifically focusing on low-energy nucleon-nucleon scattering and high-energy \Moller scattering.
We find that the non-local non-stabilizerness 
induced in both interactions is four times the anti-flatness 
(which is found to be true for any two-qubit wavefunction), 
and verify the relation between the Clifford-averaged anti-flatness and total non-stabilizerness.
For these processes, the anti-flatness is a  more experimentally accessible quantity as it can be determined from one of the final-state particles, and does not require spin correlations.
While the \Moller experiment at the Thomas Jefferson National Accelerator Facility does not include 
final-state spin measurements, the results presented here may add motivation to consider their 
future inclusion.
\end{abstract}

\maketitle
\newpage{}

\begingroup
\hypersetup{linkcolor=black}
\endgroup

\pagenumbering{gobble}

\newpage{}
\section{Introduction}
\pagenumbering{arabic}
\setcounter{page}{1}
\label{sec:intro}

\noindent
Understanding the complexity of quantum many-body systems is key to progress in an array of research and technology domains, including in accelerating future simulations using quantum computers.
To that aim, techniques quantifying the quantum complexity, and hence classical computational hardness of simulating such systems, both modest-sized and at scale, continue to improve, driven in part by challenges facing classical computation and the development of quantum computers.
One challenging area for classical simulations, in general, is predicting the properties and dynamics of matter under extreme conditions. They are plagued by sign problems, both in finding low-lying states at finite density and in their real-time evolution.
Understanding how various aspects of complexity are generated by fundamental forces, and how they evolve in few-body and many-body environments is key towards better understanding the role of quantum information in these physical phenomena and designing resource-efficient simulation strategies. 
\\

The role of entanglement -- which characterizes one aspect of quantum complexity -- has been investigated in various areas of high-energy physics (HEP) and nuclear physics (NP). For example,
building on the work of Ref.~\cite{10.21468/SciPostPhys.3.5.036,cerveralierta2019thesis} which identified maximum entanglement in Standard Model scattering processes, the suppression of fluctuations in entanglement from the S-matrix has been connected to the emergence of global symmetries beyond those explicit in the Lagrange density~\cite{Beane:2018oxh}, and further explored in Refs.~\cite{Beane:2021bpr,Low:2021ufv,Beane_2021,Liu_2023,Liu_2024}.
These are consistent with the results obtained from the large-N$_c$ limit of QCD~\cite{Kaplan_1996}, and extend to larger SU(N) spin-flavor symmetry groups~\cite{Beane:2018oxh,mcginnis2025a}, 
consistent with the results of lattice QCD calculations~\cite{Wagman:2017tmp}.
This concept has also been applied in other settings, such as the Higgs sector of
the Standard Model~\cite{carena2023a,Busoni_2025}, bosonic field theories~\cite{Chang_2024},
and higher-spin systems~\cite{hu2025a}. The maximum entanglement discussion of Ref.~\cite{10.21468/SciPostPhys.3.5.036,cerveralierta2019thesis} has been recently extended to perturbative gluon-gluon scattering in quantum chromodynamics~\cite{nunez2025a}.
Various studies of entanglement in few-nucleon scatterings~\cite{Bai:2022hfv,Bai:2023hrz,Bai:2023tey,Miller:2023ujx,Miller:2023snw,Kirchner:2023dvg,Bai:2024omg,Cavallin:2025kjn}, and in many-body nuclear systems~\cite{Johnson:2022mzk,PhysRevC.92.051303,Kruppa:2020rfa,Robin:2020aeh,Kruppa:2021yqs,Pazy:2022mmg,Tichai:2022bxr,Perez-Obiol:2023wdz,Gu:2023aoc,liu2023hints,Bulgac:2022cjg,Bulgac:2022ygo, PhysRevA.103.032426,Faba:2021kop,Faba:2022qop,Hengstenberg:2023ryt,Lacroix:2024drc,Brokemeier:2024lhq} 
have also followed.
\\

Quantum complexity beyond entanglement, in particular non-stabilizerness~\cite{gottesman1998a,Aaronson_2004}, or ``quantum magic''~\cite{Bravyi_2016}, 
is now being considered in an array of physical systems. The non-stabilizerness in a wavefunction is a measure of how much it differs from one that can be prepared efficiently classically, either as a measure of the distance to the nearest stabilizer state,  
or the number of stabilizer states needed to represent it (stabilizer rank).
As such, non-stabilizerness can provide guidance for partitioning simulations between classical and quantum hardware, both in the NISQ and fault-tolerant regimes. 
The recent development of practically-computable measures of non-stabilizerness based on Stabilizer R\'enyi Entropies (SREs)~\cite{Leone:2021rzd,Haug:2022vpg,Haug:2023hcs,Leone:2024lfr,Bittel:2025yhq} 
have triggered a wide range of works investigating non-stabilizerness in various aspects of 
few-body and many-body structure and dynamics.
In HEP and NP, this includes studies of non-stabilizerness in the dynamics of neutrino systems~\cite{Chernyshev:2024pqy}, in thermalization in gauge theories~\cite{Ebner_2025}, in simulations of lattice gauge theories~\cite{PhysRevB.111.L081102}, in high-energy particle production~\cite{PhysRevD.110.116016,CMS:2025cim} and scattering~\cite{Liu:2025qfl,Liu:2025bgw,Gargalionis:2025iqs}, in nuclear and hyper-nuclear forces~\cite{Robin:2024bdz}, in many-body nuclear models~\cite{Robin:2025wip} and atomic nuclei~\cite{Brokemeier:2024lhq}. 
These studies constitute efforts towards better understanding the role of quantum complexity in physical behaviors and designing optimal simulation strategies~\footnote{The number of non-Clifford gates required to prepare a quantum state has been shown to be lower-bounded by the non-stabilizerness in that state. However, a strict relation remains to be established~\cite{Beverland:2019jej,Leone:2021rzd}.}. 
\\

Entanglement or non-stabilizerness alone, however, do not reflect the full complexity of a quantum state.
This is because states with both large non-stabilizerness and low entanglement, and those with large entanglement and low non-stabilizerness can be prepared efficiently with classical resources. However, this becomes increasingly difficult as the entanglement or non-stabilizerness increases, respectively. Therefore it is the interplay between these two aspects that drive quantum complexity, and the need for quantum computers~\cite{Cao:2024nrx,Iannotti:2025lkb}. 
Recently it has been put forward that this interplay is captured by the non-local non-stabilizerness~\cite{Cao:2024nrx} -- the non-stabilizerness that cannot be removed from the system via local operations -- and thus large non-local non-stabilizerness will be the key feature of quantum systems that drive future quantum advantages~\cite{Beverland:2019jej,Bravyi_2016,Bravyi2019simulationofquantum,PhysRevX.6.021043,PhysRevA.71.022316,PhysRevLett.123.170502,PhysRevA.83.032317,PhysRevA.110.062427,PRXQuantum.6.020324,Gu_2024,PhysRevLett.118.090501,PRXQuantum.3.020333,PRXQuantum.2.010345,True_2022,Yoganathan_2019,10.21468/SciPostPhys.9.6.087,Bejan:2023zqm,Koh_2017,Bouland2018,zhang2024a,Qian:2025oit,Ahmadi:2022bkg,Wagner:2024jax}.
In holography theory, non-local non-stabilizerness has been shown to be responsible for gravitational back reaction and necessary for creating patterns of multipartite entanglement~\cite{Cao:2024nrx}.
Non-local non-stabilizerness has also been studied in the transverse-field Ising model \cite{Qian:2025oit}, 
and a similar concept has been studied in relation 
to the phases of matter~\cite{Korbany:2025noe}.

One issue in accessing non-local non-stabilizerness is that it requires minimization under local operations, 
which quickly becomes computationally demanding.  
On the other hand, it has recently been established that the anti-flatness of the bi-partite entanglement spectrum of a state, which is related to the total non-stabilizerness~\cite{Tirrito:2023fnw,Turkeshi:2023ctq}, can also partly capture the entanglement-non-stabilizerness interplay~\cite{Iannotti:2025lkb} and provide a lower bound to the non-local non-stabilizerness~\cite{Cao:2024nrx}.
Anti-flatness has been explored in a number of systems~\cite{cusumano2025a}, including the detection of phase transitions~\cite{sierant2025a}, in scrambling~\cite{Odavic2025a}, in the thermalization of non-Abelian gauge field theories~\cite{Ebner_2025,Ebner:2025pdm},
as well as in relation with quantum learnability~\cite{Ivaki:2025ora}.
\\

In this work, we consider the role of anti-flatness and non-local non-stabilizerness in two-body scattering processes, specifically low-energy nucleon-nucleon scattering and high-energy \Moller scattering.
While these systems do not present challenges from a computational standpoint, 
they probe the potential capabilities of these fundamental processes to drive the quantum complexity of larger many-body systems.
Previously, it has been shown that the Clauser-Horne-Shimony-Holt (CHSH) inequality can be violated in polarized \Moller scattering~\cite{PhysRevA.95.022103}, building on some of their previous works, e.g., Ref.~\cite{PhysRevA.77.012103}.  
Given the simplicity of two-particle systems, 
along with the fact that anti-flatness of the system 
can be recovered from measurements performed on one of the particles,
we propose that polarized measurements performed on 
one of the final state particles from
a set of specifically prepared polarized initial states are sufficient to reveal the changes in quantum complexity 
imparted by fundamental forces. 
Interestingly, it has recently been shown~\cite{cusumano2025a} that 
violations of the CHSH inequality depend upon the presence of local non-stabilizerness, but are suppressed by the presence of non-local non-stabilizerness.

\section{Anti-Flatness and Non-Local Non-stabilizerness}
\label{sec:AFNL}
\noindent
The most general wavefunction of a system of two spin-${1\over 2}$ particles (mapped to qubits),
$|\psi\rangle$,
requires the universal quantum gate set to prepare it from an arbitrary (classical) tensor-product state.
The vector of matrix elements of the 16 generalized Pauli operators formed from
$\overline{\sigma} \in \{ \hat X, \hat Y, \hat Z, \hat I\}$, $\hat P_{ij}=\overline{\sigma}_i\otimes \overline{\sigma}_j$, defines the $\Xi_P$ values
\begin{eqnarray}
\Xi_P & = & c_P^2/d \ ,\  
c_P\ =\ \langle\psi|\hat P|\psi\rangle
\ ,
\end{eqnarray}
where $d=2^2$ and $\sum\limits_P\Xi_P=1$.
If $|\psi\rangle$ 
is a stabilizer 
state, it
can be prepared efficiently using classical resources (using Clifford gates),
and only $d$ of the $d^2$ quantities $\Xi_P$ are non-zero, 
with values  $\Xi_P=1/d$, 
such that
$\xi=d \sum\limits_P\Xi_P^2=1$, along with other analogous relations.
For a non-stabilizer state, while $\sum\limits_P\Xi_P=1$ remains valid,
$1/d\le \xi<1$.   
A family of SREs, ${\cal M}_\alpha$ are defined by~\cite{Leone:2021rzd}
\begin{eqnarray}
{\cal M}_\alpha (|\psi\rangle) 
& = & 
{1\over 1-\alpha} \log_2\sum_P \Xi_P^\alpha \ -\  \log_2 d \; .
\label{eq:magicalpha}
\end{eqnarray}
Out of these measures the stabilizer 2-R\'enyi entropy $\mathcal{M}_2(|\psi\rangle)$ has been shown to satisfy property of monoticity~\cite{Leone:2024lfr,Haug:2023hcs}. Here we will pursue ${\cal M}_{\rm lin}(|\psi\rangle)$, the linear version of $\mathcal{M}_2$, 
\begin{eqnarray}
{\cal M}_{\rm lin}(|\psi\rangle) \ = \ 1-\xi \; .
\label{eq:magic_lin}
\end{eqnarray}
The main reason behind this choice is that ${\cal M}_{\rm lin}(|\psi\rangle)$ has been shown to be directly related to some aspect of interplay between entanglement and non-stabilizerness~\cite{Tirrito:2023fnw}, 
which will become clearer below. Moreover, 
${\cal M}_{\rm lin}(|\psi\rangle)$ has been shown to be a strong monotone~\cite{Leone:2024lfr} and directly yields the stabilizer 2-R\'enyi entropy through
\begin{eqnarray}
{\cal M}_2 (|\psi\rangle) 
& = & 
-\log_2 \xi
\ =\ 
-\log_2 \left(1-{\cal M}_{\rm lin}(|\psi\rangle) \right) \; .
\end{eqnarray}
As defined, these measures of non-stabilizerness contain contributions from both local and non-local non-stabilizerness.
Given that the local non-stabilizerness can be eliminated, by definition, by applications of local unitary operators, the bi-partite non-local non-stabilizerness
\footnote{The definition of non-local non-stabilizerness can be straightforwardly generalized to systems
with more than two qubits, both for bi-partitions and multi-partitions~\cite{cao2024b,Cao:2024nrx}.
}
is defined by~\cite{cao2024b,Cao:2024nrx}
\begin{eqnarray}
{\cal M}_i^{(NL)} (|\psi\rangle) 
& = 
\min_{\hat{U}_A \otimes \hat{U}_B} \ \left[  {\cal M}_i^{(NL)} \left(\hat U_A\otimes\hat U_B |\psi\rangle\right) \right] \; , \nonumber \\
& \hspace{4cm} i \equiv \alpha, {\rm lin}
\ ,
\label{eq:NLM}
\end{eqnarray}
where $A$ and $B$ denote the two chosen partitions (here the two qubits) and
where the minimization is over all possible local unitary transformations on $A$ and $B$, providing a basis-independent quantity.
For  the two-qubit system,  the minimization  corresponds to the simple task of searching over the six angles defining transformations over each Bloch sphere.  In some instances, this can be accomplished analytically, while in many others it requires a numerical procedure.
\\

It has been recently shown that measures of total non-stabilizerness
and non-local non-stabilizerness in a pure state $\ket{\psi}$ can be related to the anti-flatness of the bi-partite entanglement spectrum of that state.
Such anti-flatness is defined by a polynomial function of the reduced density matrix $\hat\rho_A={\rm Tr}_B\ \hat\rho_{AB}$ of subsystem A~\cite{Tirrito:2023fnw} as
\begin{eqnarray}
{\cal F}_A (\ket{\psi}) & = & \langle\rho_A^2\rangle - \langle\rho_A\rangle^2
\ =\ 
{\rm Tr} (\rho_A^3) - \left({\rm Tr} \rho_A^2\right)^2 \; ,
\label{eq:Antiflat}
\end{eqnarray}
and corresponds to the variance of the reduced density matrix. 
The two contributions to the anti-flatness cancel when the non-zero eigenvalues of $\hat\rho_A$ are equal (flat entanglement spectrum), which occurs if $\ket{\psi}$ is unentangled or has no non-stabilizerness~\cite{Tirrito:2023fnw}.
It is important to note that, similarly to the non-local non-stabilizerness, the anti-flatness is independent on the local basis, but is not invariant under global Clifford operations.
\\

In Ref.~\cite{Tirrito:2023fnw} it has been shown that the total linear non-stabilizerness is directly proportional to the 
average anti-flatness associated with applications of all possible combinations of Clifford gates to the complete state:
\begin{equation}
    \langle \mathcal{F}_A (\hat{\Gamma} \ket{\psi}) \rangle_\mathcal{C} 
    = c(d,d_A) \ \mathcal{M}_{\rm lin} (\ket{\psi}) \; ,
\label{eq:Cliffav_AF}
\end{equation}
where the left-hand side is the anti-flatness of $\hat{\Gamma} \ket{\psi}$ averaged over 
Clifford unitaries $\hat{\Gamma} \in \mathcal{C}$, and the proportionality constant
$c(d,d_A)$ is given by 
\begin{equation}
    c(d,d_A) = \frac{(d^2 - d_A^2) (d_A^2 -1)}{(d^2-1) (d+2) \, d_A^2} 
    \; ,
    \label{eq:cfun}
\end{equation}
where $d_A$ is the dimensionality of system-A Hilbert space. 
The averaging over Clifford operators provides a quantity that is Clifford-invariant. At the same time, it re-distributes the non-stabilizerness among local and non-local components, and thus can re-introduce local non-stabilizerness.
\\

Further, Ref.~\cite{Cao:2024nrx} showed that the presence of non-local non-stabilizerness is a necessary and sufficient condition for anti-flatness of the entanglement spectrum
\footnote{The sufficiency condition is verified for most non-stabilizerness measures, except for some particular non-integer SREs~\cite{Cao:2024nrx}.}, 
and is lower bounded by the latter.
In the present study, we find that the anti-flatness in two-particle scattering (mapped onto two qubits) is exactly proportional to the non-local linear non-stabilizerness.

\section{Two-Particle Scattering}
\label{sec:TwoParts}
\noindent
As the dominant force between particles in nuclei and many processes involving fundamental particles, 
two-particle processes encapsulated by the S-matrix and T-matrix provide 
the basic building blocks of quantum complexity in many-body systems.
For demonstrative purposes, motivated by what might be experimentally possible,
we examine the anti-flatness and non-local non-stabilizerness in 
the spin-sector of 
low-energy (non-relativistic) S-wave nucleon-nucleon scattering, 
and high-energy (relativistic) \Moller scattering, 
$e^-e^-\rightarrow e^-e^-$.
These processes probe different sectors of the fundamental forces of Nature. While the simple frameworks adopted in this work yield similar analytic structures for both processes, they already reveal notable differences in the ability of the underlying forces to generate and remove non-local non-stabilizerness and entanglement.
\\

\subsection{Nucleon-Nucleon Scattering}
\label{sec:NN}
\noindent
Low-energy $S$-wave 
neutron-proton ($np$)
scattering is described by an S-matrix of the form
\begin{eqnarray}
\hat S & = & 
\left(
\begin{array}{cccc}
e^{2 i \delta_1} & 0 & 0 & 0  \\
0 & {{1\over 2}\left(e^{2 i \delta_1} + e^{2 i \delta_0}\right)}
& {{1\over 2}\left(e^{2 i \delta_1} - e^{2 i \delta_0}\right)} & 0\\
0 & {{1\over 2}\left(e^{2 i \delta_1} - e^{2 i \delta_0}\right)}
& {{1\over 2}\left(e^{2 i \delta_1} + e^{2 i \delta_0}\right)} & 0\\
0 & 0 & 0  &e^{2 i \delta_1} 
\end{array}
\right)
\ , \nonumber \\
\label{eq:NNSmat}
\end{eqnarray}
where $\delta_{0,1}$ are the phase shifts in the $S=0$ and $S=1$ spin channels.
These are the dominant, but not only, channels contributing at low-energies, and a more complete analysis would include contributions from P-wave, SD-mixing (induced by the tensor force) and higher.

When applied to some initial incoming state, the S-matrix in Eq.~(\ref{eq:NNSmat}) defines the asymptotic final state of the scattering system.
In Ref.~\cite{Robin:2024bdz}, we studied the non-stabilizing power of the scattering S-matrix as the average SRE induced by the S-matrix acting on the set of the 60 two-qubit stabilizer states (listed in App.~\ref{app:stabs}):
\begin{align}
    \overline{\mathcal{M}_{\rm lin}}(\hat {\bf S}) \equiv \frac{1}{60} \sum_{i=1}^{60}  \mathcal{M}_{\rm lin}\left( \hat {\bf S} \ket{\psi_i} \right) \; .
\label{eq:Magic_Power}
\end{align}
This quantity included both the local and non-local non-stabilizing powers, which were not separated.
In order to separate them, we define here the non-local non-stabilizing power and anti-flatness power in a similar manner.
To further isolate the induced changes of entanglement, we however choose to restrict the average to the 36 tensor-product states (formed from the 6 single-qubit stabilizer states) only. That is,
\begin{align}
   \overline{\overline{\mathcal{M}_{\rm lin}^{(NL)}}}(\hat {\bf S}) &\equiv \frac{1}{36} \sum_{i=1}^{36}  \mathcal{M}^{(NL)}_{\rm lin} \left( \hat {\bf S} \ket{\psi_i} \right) \; , 
   \label{eq:NL_Magic_Power}\\
    \overline{\overline{\mathcal{F}_A}}(\hat {\bf S}) &\equiv \frac{1}{36} \sum_{i=1}^{36}  \mathcal{F}_A \left( \hat {\bf S} \ket{\psi_i} \right) \; .
\label{eq:AF_Power}
\end{align}
Quantities averaged over the full set of stabilizer states are denoted with a single bar $\overline{O}$, while those averaged over the tensor-product stabilizer states are denoted with a double bar $\overline{\overline{O}}$.
\\

\begin{figure}[!ht]
    \centering
    \includegraphics[width=\columnwidth]{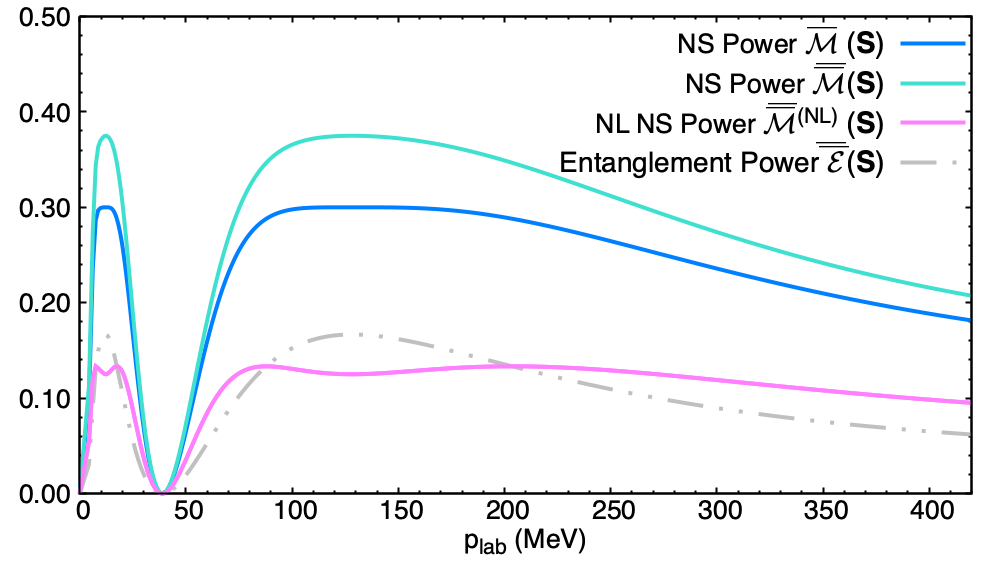}
    \caption{The total non-stabilizing (NS) powers, $\overline{{\cal M}_{\rm lin}}(\hat S)$ and $\overline{\overline{{\cal M}_{\rm lin}}}(\hat S)$, 
    and the non-local non-stabilizing (NL NS) power, $\overline{\overline{{\cal M}_{\rm lin}^{(NL)}}}(\hat S)$, 
    for low-energy S-wave 
    neutron-proton
    scattering
    as a function of momentum in the laboratory.
    The difference in phase shifts are determined from the Nijmegen phase-shift analysis~\cite{PhysRevC.49.2950,NNonline}.
    The entanglement power is also shown, as a dashed line.
    }
    \label{fig:NNnlm}
\end{figure}

Direct optimization over the 6 Euler angles defining the two Bloch spheres yields the following expression for the non-local non-stabilizing power
%
\begin{align}
 \overline{\overline{{\cal M}_{\rm lin}^{(NL)}}} (\hat S)  
=
{1\over 48}\left( 11 + 5 \cos\left(4 \Delta\delta\right)\right) \sin^2\left(2 \Delta\delta\right) \; ,
\label{eq:NLMagNN}
\end{align}
where $\Delta \delta = \delta_1 - \delta_0$.
This is to be compared to the total non-stabilizing power, 
which is given by~\cite{Robin:2024bdz}
\begin{eqnarray}
 \overline{{\cal M}_{\rm lin}} (\hat S)   & = & 
{3\over 20}\left( 3 +  \cos\left(4 \Delta\delta\right)\right) \sin^2\left(2 \Delta\delta\right) 
\; .
\label{eq:MagNN}
\end{eqnarray}
The results obtained using Nijmegen Nijm93 phase shifts~\cite{PhysRevC.49.2950,NNonline} 
(determined from fitting NN experimental data)
are shown in Fig.~\ref{fig:NNnlm}. 
For comparison, 
we also show the linear non-stabilizing power $\overline{\overline{{\cal M}_{\rm lin}}} (\hat S)$ averaged over tensor-product stabilizer states only,
along with
the entanglement power for information \footnote{
The corresponding entanglement power is given by~\cite{Beane:2018oxh,Robin:2024bdz}
\begin{eqnarray}
 \overline{\overline{\cal E}} (\hat S)   & = & 
{1\over 6} \sin^2\left(2 \Delta\delta\right) 
\; .
\label{eq:ESNN}
\end{eqnarray}
}.
It is seen that in the regime $p_{lab} \lesssim 150 $ MeV, 
only about one
third of the non-stabilizerness generated is non-local, while the rest can be eliminated via local basis transformation.
\\

Further, we find that the non-local non-stabilizing power is exactly equal to the anti-flatness power, 
{\it i.e.}  $4 \, \overline{\overline{{\cal F}_A}} (\hat S) = \overline{\overline{{\cal M}_{\rm lin}^{(NL)}}} (\hat S)$.
In fact, we find this exact equality to be general for 2-qubit states $\ket{\psi}$, as
\begin{align}
    4\ \mathcal{F}_A (\ket{\psi}) = \mathcal{M}_{\rm lin}^{(NL)}(\ket{\psi}) \; .
\label{eq:AF_NLmagic_2qubit}
\end{align}
We have verified this result numerically for a large set ($10^5$)
of random 2-qubit states, with precision $\sim 10^{-28}$. 
This equality between non-local linear non-stabilizerness and anti-flatness is however only valid for 2-qubit systems. 
For $n$-qubit systems with $n>2$, the anti-flatness only provides a lower bound on the non-local non-stabilizerness, which contains higher-order effects. 
It was also recently shown (after submission of the present paper) that the equality in Eq.~\eqref{eq:AF_NLmagic_2qubit} does not hold for higher-dimensional 2-qudit systems, such as 2-qutrits~\cite{Busoni:2026lvp}.

We have also numerically verified the relation in Eq.~\eqref{eq:Cliffav_AF}, 
$\langle \mathcal{F}_A (\Gamma \ket{\psi}) \rangle_\mathcal{C} = c(d,d_A) \ \mathcal{M}_{\rm lin} (\ket{\psi})$.
\\

We analyze in more detail the individual contribution of outgoing states produced from different groups of initial stabilizer states.
As we identified previously~\cite{Robin:2024bdz}, these 36 states divide into 3 groups,
Group-1, 2, 3, in which there are 6, 6, 24 tensor-product initial states, respectively. 
These groups are chosen so that the states in each group yield the same total outgoing non-stabilizerness, 
and are listed in App.~\ref{app:np}.
The anti-flatness is again equal to four times the non-local linear non-stabilizerness, and interestingly, we find that, in each group, the non-local linear non-stabilizerness is also proportional to the total linear non-stabilizerness:
\begin{eqnarray}
{\rm Group-1} & : & {\cal F}_A  = {\cal M}_{\rm lin}^{(NL)}\ =\ {\cal M}_{\rm lin}\ =\ 0\ , \label{eq:NNQC1} 
\\
{\rm Group-2} & : & 4\ {\cal F}_A  = {\cal M}_{\rm lin}^{(NL)}\ =\ {\cal M}_{\rm lin}\ =\ {1\over 4}\sin^2\left(4 \Delta\delta\right) ,  \nonumber \\
\label{eq:NNQC2}
\\
{\rm Group-3} & : & 4\ {\cal F}_A  = {\cal M}_{\rm lin}^{(NL)}\ =\ {1\over 3}{\cal M}_{\rm lin} \nonumber \\
&& \hspace{1cm} = {1\over 32}\left( 7 + \cos\left(4 \Delta\delta\right) \right) \sin^2\left(2 \Delta\delta\right) 
 .
\label{eq:NNQCs}
\end{eqnarray}
The proportionality constant is however different for each of the groups.
In contrast to Groups-2,3, scattering of Group-1 states does not create non-stabilizerness or anti-flatness.
For scattering of states in Group-2, the total non-stabilizerness of the final states is saturated by the non-local non-stabilizerness.  That is to say that all the non-stabilizerness generated is bound to the outgoing system and there is no non-stabilizerness in the final state that can be removed by changes of local basis. In Group-3, the non-local non-stabilizerness constitutes only a third of the total non-stabilizerness.
In App.~\ref{app:np}, we show 
the non-stabilizerness, non-local non-stabilizerness (here equivalent to anti-flatness) and entanglement for scattering of states in each group, and also show results when considering entangled stabilizer states.
\\

Tensor-product stabilizer states representative of each group, 
denoted by $|\psi_{1,2,3}\rangle$, that are candidates for experimental initial-state preparation are,
\begin{eqnarray}
&& |\psi_1\rangle \ =\ |\uparrow\rangle \otimes |\uparrow\rangle 
\ \ ,\ \ 
 |\psi_2\rangle \ =\ |\uparrow\rangle \otimes |\downarrow\rangle  
 \ \ , \ \ \nonumber \\
 && |\psi_3\rangle \ =\ 
{1\over\sqrt{2}} \left[\ |\uparrow\rangle + |\downarrow\rangle \ \right]\otimes |\uparrow\rangle 
 \ ,
 \label{eq:NNstabs}
\end{eqnarray}
corresponding to states 33, 34 and 25 in Table.~\ref{tab:TwoQstabs}.
Experimentally, the wavefunction associated with the selected Group-3 state can be prepared by a single-spin $\pi/2$ rotation about the y-axis. \\

\begin{figure}[!ht]
    \centering
    \includegraphics[width=\columnwidth]{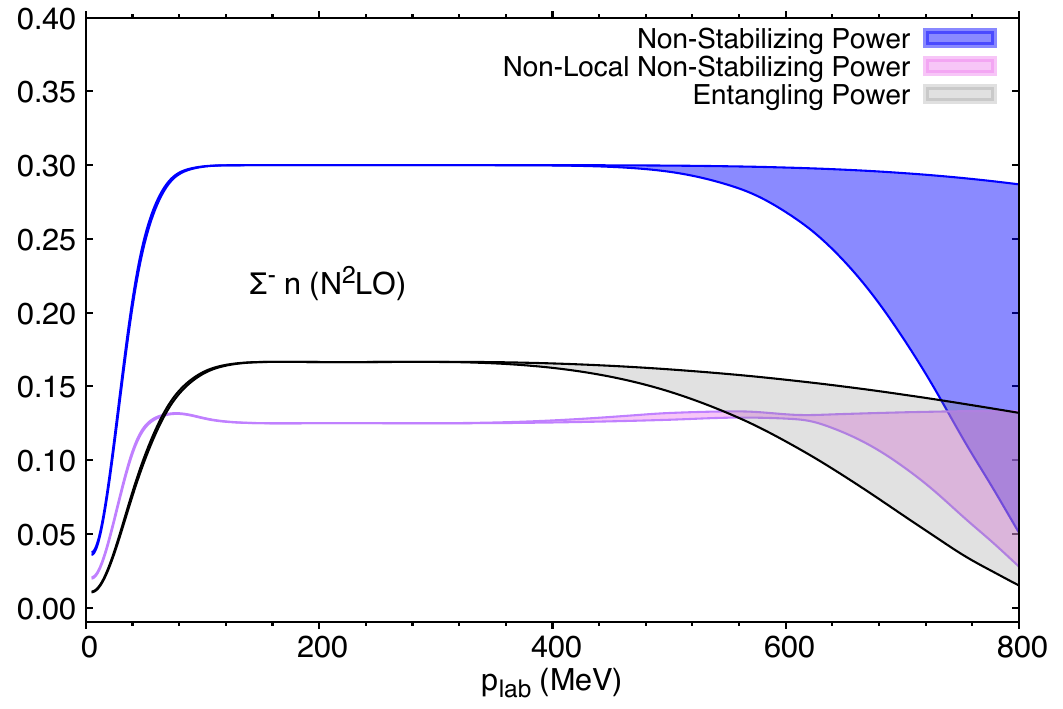}
    \caption{Total non-stabilizing power, $\overline{{\cal M}}_{\rm lin}(\hat S)$, 
    non-local non-stabilizing power, $\overline{\overline{{\cal M}_{\rm lin}^{(NL)}}}(\hat S)$, 
    and entanglement power,  $\overline{{\cal E}_{\rm lin}}(\hat S)$,
    for low-energy S-wave $\Sigma^- n$ scattering
    as a function of momentum in the laboratory.
    The results are obtained using N$^2$LO chiral effective field theory phase shifts from Ref.~\cite{Haidenbauer:2023qhf}.
    We have assumed isospin symmetry between $\Sigma^+p$ and $\Sigma^-n$, and neglected Coulomb interactions.
    The uncertainty bands represent the maximum and minimum values in non-stabilizerness and entanglement derived from the N$^2$LO phase-shift uncertainty bands~\cite{Haidenbauer:2023qhf}.
    }
    \label{fig:Sign}
\end{figure}

We have previously found that the entanglement and (total) non-stabilizing power in $\Sigma^- n$ scattering 
is large and energy independent over a large range, after a rapid rise at low-energies~\cite{Robin:2024bdz}.
Extending the above analysis to the  $\Sigma^- n$ sector, shows that the non-local non-stabilizerness (or anti-flatness) exhibit similar behaviors.  They rise rapidly and remain constant, with a small amplitude modulation over an extended energy interval. 
Non-local non-stabilizerness is found to capture less than half ($\simeq 0.4$) of the total non-stabilizerness, and this fraction remains constant even at large momenta. 
This is shown in Fig.~\ref{fig:Sign}.
%

\subsection{High-Energy \Moller Scattering}
\label{sec:HEM}
\noindent
High-energy \Moller scattering provides a sensitive probe for new physics, and establishes increasingly precise constraints on electroweak interactions in 
the Standard Model.  For example, the  MOLLER experiment at the Thomas Jefferson National Accelerator Facility~\cite{MOLLER:2014iki,MollerJLab} is designed to measure the parity-violating asymmetry, $A_{PV}$, 
to a precision of $0.7$ ppb, 
thereby probing the interference of 
amplitude from the exchange of a neutral weak-gauge boson, $Z^0$, and the electromagnetic amplitude with remarkable sensitivity.
This process also provides somewhat of a complementary process to NN scattering for 
exposing non-local non-stabilizerness and anti-flatness in two-body scattering. 
It has been considered in previous works exploring quantum complexity in fundamental processes, 
including (the first) work that highlighted that entanglement seems to be maximal~\cite{10.21468/SciPostPhys.3.5.036,cerveralierta2019thesis}, and a recent consideration of total non-stabilizerness~\cite{Liu:2025qfl}.
Following Ref.~\cite{Liu:2025qfl}, instead of considering the structure of the 
complete S-matrix, we   consider the final state wavefunction generated 
at leading order in perturbation theory 
(in the electromagnetic coupling constant, $e$)
by a single insertion of the T-matrix.  
This preserves unitarity up to ${\cal O}(e^4)$, and necessarily requires a projective measurement on the two-electron final state.

The post-scattering wavefunction can be determined from the expressions in App.~\ref{app:Moll}, and using the helicity-amplitude basis, the scattering amplitude matrix 
in the high-energy limit ($m_e\rightarrow 0$)
becomes~\cite{10.21468/SciPostPhys.3.5.036,cerveralierta2019thesis,blasone2024a}, 
\begin{eqnarray}
{\cal A} & \propto & 
\left(
\begin{array}{cccc}
 -8 \csc ^2(\theta ) & 0 & 0 & 0 \\
 0 & -2 \cot ^2\left(\frac{\theta }{2}\right) & 2 \tan ^2\left(\frac{\theta
   }{2}\right) & 0 \\
 0 & 2 \tan ^2\left(\frac{\theta }{2}\right) & -2 \cot ^2\left(\frac{\theta
   }{2}\right) & 0 \\
 0 & 0 & 0 & -8 \csc ^2(\theta ) \\
\end{array}
\right)
 , \nonumber \\
\label{eq:MollAmp}
\end{eqnarray}
where $\theta$ is the center-of-momentum scattering angle.
The proportionality constant, including factors of $e^2$ are inconsequential upon renormalizing the final state wavefunction.
For any initial state wavefunction, $|\psi\rangle$, 
the final state wavefunction is given by 
$|\chi\rangle = \mathcal{N}\ {\cal A}|\psi\rangle$.
The normalization constant, $\mathcal{N}$, is determined (up to an overall phase) 
by $\langle\chi|\chi\rangle=1$.
In contrast to Ref.~\cite{Liu:2025qfl} with amplitudes given in terms of the computational basis (because of the basis dependence of the total non-stabilizerness), we use the helicity basis to write ${\cal A}$, but any single-particle basis would have served our purposes.
The total non-stabilizerness, non-local non-stabilizerness and anti-flatness powers take the same expression as in Eqs.~\eqref{eq:Magic_Power}, \eqref{eq:NL_Magic_Power} and~\eqref{eq:AF_Power}, with $\hat S \ket{\psi_i} \rightarrow \mathcal{N}\ \hat{\cal A}|\psi_i\rangle$.
\\

There are five different groups of  stabilizer states with respect to the total non-stabilizerness injected by the scattering amplitude in Eq.~(\ref{eq:MollAmp}), which are given in 
Eq.~(\ref{eq:GroupsMoll}) in App.~\ref{app:Moll}.  Each state in a given group gives rise to a final state with the same total non-stabilizerness.
Representative initial tensor-product states from each group are 
\begin{eqnarray}
&& |\psi_1\rangle \ =\ |\uparrow\rangle \otimes |\uparrow\rangle 
\ \ ,\ \ \nonumber \\
&& |\psi_2\rangle \ =\ 
 {1\over\sqrt{2}} \left[\ |\uparrow\rangle + |\downarrow\rangle \ \right]\otimes 
 {1\over\sqrt{2}} \left[\ |\uparrow\rangle + |\downarrow\rangle \ \right]
 \ \ , \ \ \nonumber \\
&& |\psi_3\rangle \ =\ 
 {1\over\sqrt{2}} \left[\ |\uparrow\rangle + |\downarrow\rangle \ \right]\otimes 
 {1\over\sqrt{2}} \left[\ |\uparrow\rangle - |\downarrow\rangle \ \right]
 \ \ , 
 \nonumber\\
&&  |\psi_4\rangle \ =\ 
|\uparrow\rangle \otimes |\downarrow\rangle 
 \ \ ,\ \ \nonumber \\
&& |\psi_{5a}\rangle \ =\ 
 {1\over\sqrt{2}} \left[\ |\uparrow\rangle + i |\downarrow\rangle \ \right]\otimes 
 {1\over\sqrt{2}} \left[\ |\uparrow\rangle + |\downarrow\rangle \ \right]
 \ \ ,\ \ \nonumber \\
&& |\psi_{5b}\rangle \ =\ 
|\uparrow\rangle \otimes 
 {1\over\sqrt{2}} \left[\ |\uparrow\rangle + |\downarrow\rangle \ \right]
 \ .
 \label{eq:Mollerinitialstates}
\end{eqnarray}

\begin{figure*}[!ht]
    \centering
    \includegraphics[width=0.8\textwidth]{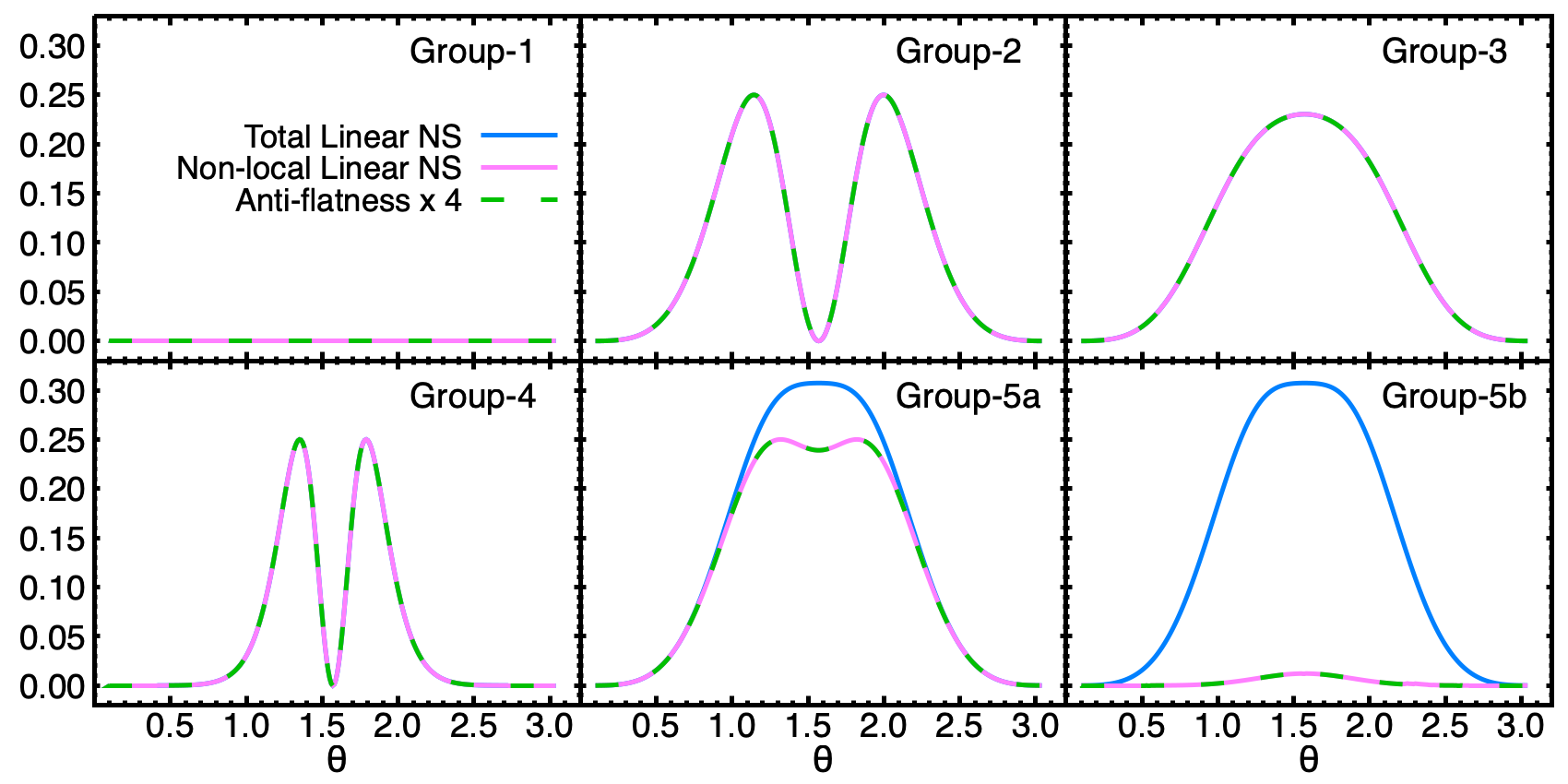}
    \caption{
    The linear non-stabilizerness (NS), ${\cal M}_{\rm lin}(\ket{\chi_i})$, the non-local linear non-stabilizerness (NL NS), 
    ${\cal M}_{\rm lin}^{(NL)}(\ket{\chi_i})$ and the anti-flatness, ${\cal F}_A(\ket{\chi_i})$ (multiplied by a factor 4), for outgoing states 
    $\ket{\chi_i} = \mathcal{N} \hat{\cal A} \ket{\psi_i}$, corresponding to
    the distinct groups of initial-state helicity wavefunctions $\ket{\psi_i}$
    for high-energy \Moller scattering.
    The upper panels from left to right show the results for tensor-product states from Group-1, 2, 3,
    while the lower panels show the results for Group-4, 5a, 5b.
    The green curves associated with the ${\cal M}_{\rm lin}$ are from 
    Eq.~(\ref{eq:MollerM2ana}),
    while the dashed blue curves associated with the ${\cal F}_A$ are from 
    Eq.~(\ref{eq:MollerAF}).
    The pink curves are from numerical minimization of ${\cal M}_{\rm lin}^{(NL)}(\ket{\chi_i})$, 
    using Eq.~(\ref{eq:NLM}).
    }
    \label{fig:MOLLHECOMPgr1to5b}
\end{figure*}

The values of the total linear non-stabilizerness, 
${\cal M}_{\rm lin}(\ket{\chi_i})$,
defined in Eq.~(\ref{eq:magic_lin}),
of the associated 
wavefunction
$\ket{\chi_i} = \mathcal{N} \hat{\mathcal{A}} \ket{\psi_i}$
for each of the groups, 
are
\begin{eqnarray}
&& G1\ : \    0
\ ,\nonumber\\
&& G2\ : \  
\frac{64 \sin ^4(\theta ) \cos ^2(\theta )}{(\cos (2 \theta )+3)^4}
\ ,\nonumber\\
&& G3\ : \    
\frac{1024 \sin ^4(\theta ) (20 \cos (2 \theta )+\cos (4 \theta )+43)^2}{(12
   \cos (2 \theta )+\cos (4 \theta )+51)^4}
\ , \nonumber \\
&& G4\ : \   
\frac{4 \cot ^8\left(\frac{\theta }{2}\right) \left(\cot ^8\left(\frac{\theta
   }{2}\right)-1\right)^2}{\left(\cot ^8\left(\frac{\theta
   }{2}\right)+1\right)^4}
   \ , \\
&& G5\ : \ \nonumber \\
&& \frac{32 \sin ^4(\theta ) (799 \cos (2 \theta )-10 \cos (4 \theta )+\cos (6
   \theta )+1258)}{(\cos (2 \theta )+7)^6}
   \ . \nonumber
 \label{eq:MollerM2ana}
\end{eqnarray}
The corresponding anti-flatness is found to recover the results of numerical minimization of 
${\cal M}_{\rm lin}^{(NL)} = 4\  {\cal F}_A $ for each group, which are found to be
\begin{eqnarray}
&& G1\ :  
{\cal M}_{\rm lin}\ =\ 0
\ ,\nonumber\\
&& G2\ :  
 {\cal M}_{\rm lin}
\ ,\nonumber\\
&& G3\ :  
 {\cal M}_{\rm lin}
\ ,\nonumber\\
&& G4\ :  
 {\cal M}_{\rm lin}
\ ,\nonumber\\
&& G5a\ :  
 \frac{256 \sin ^4(\theta )}{(\cos (2 \theta )+7)^8}
\nonumber \\
&& \times  (\cos (2 \theta )+15)^2 (28 \cos (2 \theta )+\cos(4 \theta )+35)
   \leq {\cal M}_{\rm lin}\ ,\nonumber\\
&& G5b\ :  
 \frac{128 \sin ^8(\theta )}{(\cos (2 \theta )+7)^8} \nonumber \\ 
&& \times  (175 \cos (2 \theta )+18 \cos (4 \theta )+\cos (6
   \theta )+318)
   \ll {\cal M}_{\rm lin} 
   \ . \nonumber \\
\label{eq:MollerAF}
\end{eqnarray}
Notice that the states in Group-5 (G5)
produce final states that have two different values of anti-flatness.
Therefore, we further sub-divide these states into Group-5a and Group-5b,
as given in  Eq.~(\ref{eq:GroupsMollGrp3ab}) in App.~\ref{app:Moll}.
For all states, the relation $\mathcal{M}_{\rm lin}^{(NL)}(\ket{\chi_i}) = 4\, \mathcal{F}_A(\ket{\chi_i})$ is verified. 
For Group-1,2,3,4, the outgoing states only exhibit non-local non-stabilizerness as $\mathcal{M}_{\rm lin}^{(NL)}(\ket{\chi_i}) = \mathcal{M}_{\rm lin}(\ket{\chi_i})$. This is not the case for outgoing states generated from stabilizer states from Group-5a and Group-5a, which also display some local non-stabilizerness.
In finding the non-local non-stabilizerness, 
we are unable to arrive at closed-from results for ${\cal M}_{\rm lin}^{(NL)}$ by analytic minimization of local unitary transformations for some of the groups.  
However, numerical minimization over the two Bloch spheres is found to be effective.
The resulting non-local non-stabilizerness, as well as the total non-stabilizerness and anti-flatness, for initial tensor product states in each of the groups,
are shown in Fig.~\ref{fig:MOLLHECOMPgr1to5b}.
For completeness, the total linear non-stabilizerness ${\cal M}_{\rm lin}(\ket{\chi_i})$, linear entanglement entropy, and non-local non-stabilizerness ${\cal M}_{\rm lin}^{(NL)}(\ket{\chi_i})$ of the outgoing states $\ket{\chi_i} = \mathcal{N} \ket{\psi_i}$ for each groups, including 
for entangled initial states, are shown in  Fig.~\ref{fig:Moller_2x3} in App.~\ref{app:Moll}.
\\

\begin{figure}[!ht]
    \centering
    \includegraphics[width=\columnwidth]{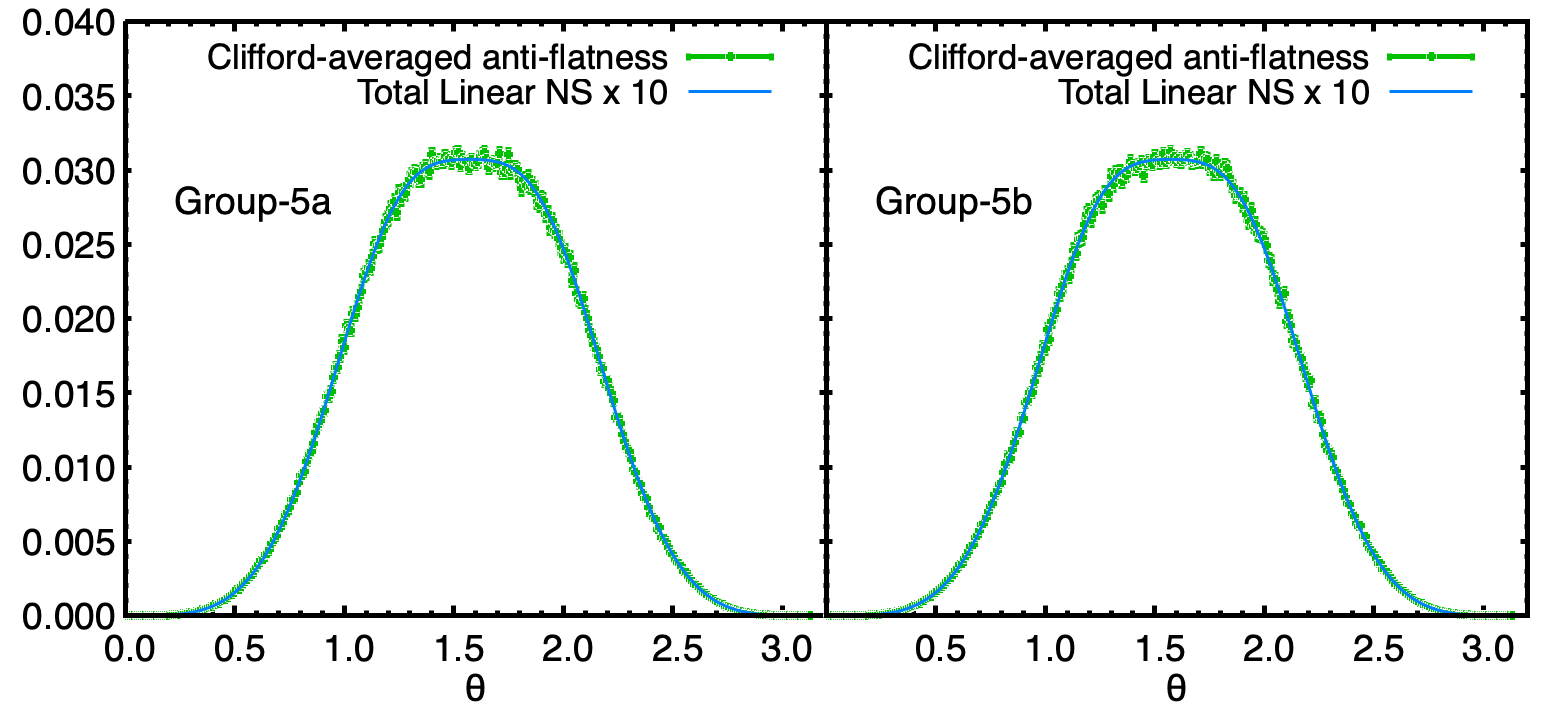}
    \caption{The Clifford-averaged anti-flatness, $\langle \mathcal{F}_A (\Gamma \ket{\chi_i}) \rangle_\mathcal{C}$,
    (green points) compared with   
    $ c(d,d_A) {\mathcal{M}}_{\rm lin}(\ket{\chi_i})$ (blue curve) in \Moller scattering
    from initial states
    $|\psi_{5a}\rangle $ (left panel)  and  $|\psi_{5b}\rangle $ (right panel)
    in Eq.~(\ref{eq:Mollerinitialstates}).
    The un-averaged values of anti-flatness are shown in Fig.~\ref{fig:MOLLHECOMPgr1to5b}.
    }
    \label{fig:Moller_AFavMlinGRP5}
\end{figure}

The relation between the anti-flatness averaged over the Clifford group,
given in Eq.~(\ref{eq:Cliffav_AF}) has been numerically verified. 
Figure~\ref{fig:Moller_AFavMlinGRP5} shows 
the Clifford-averaged anti-flatness and the total linear non-stabilizerness divided by 10
for the states in Group-5a and 5b.
The averaged results are consistent with the relation
$\langle \mathcal{F}_A (\Gamma \ket{\psi}) \rangle_\mathcal{C} = 
c(d,d_A) \ \mathcal{M}_{\rm lin} (\ket{\psi})$ 
in Eq.~(\ref{eq:Cliffav_AF}),
where $c(d,d_A)$ in Eq.~(\ref{eq:cfun}) is
$c(4,2)=1/10$.
The estimators of $\langle \mathcal{F}_A (\Gamma \ket{\psi}) \rangle_\mathcal{C}$ 
in Fig.~\ref{fig:Moller_AFavMlinGRP5} at each angle
were 
determined from the mean 
and standard deviation of an ensemble of $\mathcal{F}_A$ generated from 
$5\times 10^3$ samples of random Clifford gates applied to the final state wavefunction.
\\

The current \Moller scattering experiment (MOLLER) at JLab~\cite{MOLLER:2014iki,MollerJLab} 
is able to prepare arbitrary polarized initial states with high precision, 
but currently lacks the capability to measure spin components of final state particles.
This means that the anti-flatness cannot be measured with this generation of experiment. 
However, with sufficient motivation,
there could be potential for including
such capabilities in a next-generation MOLLER program.

\subsection{Non-local Non-stabilizerness Generation from Entangled Initial States: Nuclear Force Versus  
Quantum Electrodynamics}
\noindent
We have so far focused on the non-local non-stabilizerness and anti-flatness produced in scattering processes in which the two particles are initially unentangled.
Since non-local non-stabilizerness requires entanglement, this allowed us to isolate the 
effectiveness
of the underlying interaction to generate both the needed entanglement and the corresponding non-local non-stabilizerness.
This is also justified from an experimental point of view, as in laboratory scattering experiments the initial incoming particles are typically unentangled.

Nevertheless it is also interesting to investigate the effect of scattering when the two initial particles are prepared in an entangled stabilizer state. 
These states are those numbered 37 to 60 in Table~\ref{tab:TwoQstabs} in App.~\ref{app:stabs}. They are maximally entangled but possess, by definition, no non-stabilizerness (local or non-local) and no anti-flatness.
Thus studying scattering from these states can tell us about whether the underlying interaction can build on the initial entanglement to generate anti-flatness (complex entanglement patterns), and non-local non-stabilizerness. For example, is the non-local non-stabilizerness enhanced when the initial states are entangled?

\begin{figure*}[!ht]
    \centering
    \includegraphics[width=0.8\textwidth]{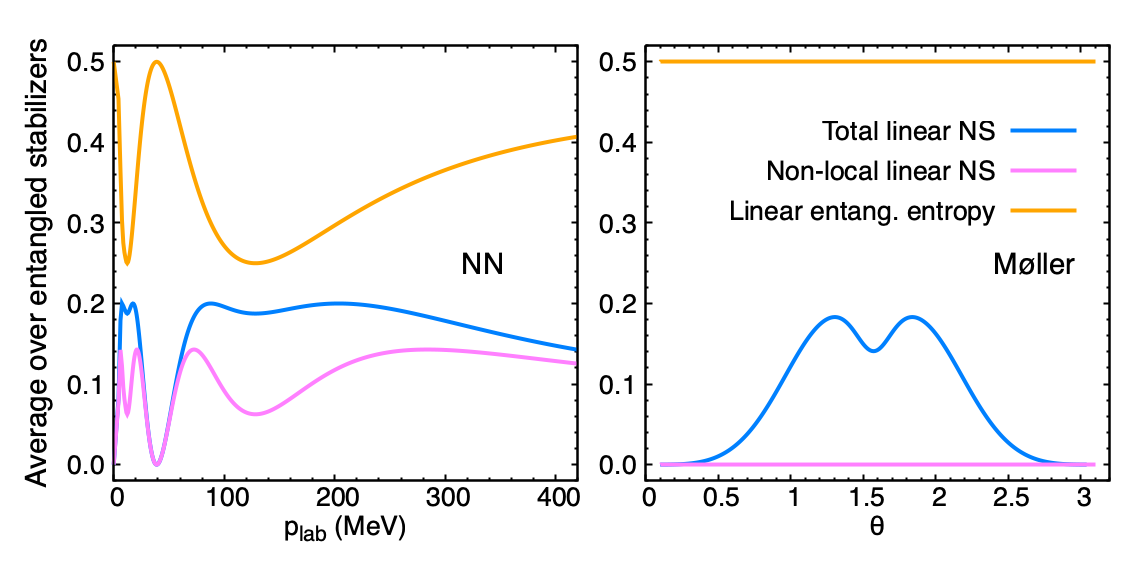}
    \caption{The total linear non-stabilizerness (NS), non-local linear non-stabilizerness (NL NS) (anti-flatness $\times$ 4), and linear 
    entanglement entropy in NN scattering and \Moller scattering final states, averaged over  
     all 24 entangled stabilizer states. }
    \label{fig:np_Moll_NLmag_entang}
\end{figure*}

Figure~\ref{fig:np_Moll_NLmag_entang} shows the final-state total non-stabilizerness, non-local non-stabilizerness (or anti-flatness) and linear entanglement entropy averaged over initial entangled stabilizer states
for both NN and \Moller scattering.~\footnote{The difference between the average final-state linear entanglement entropy and its maximal value of $1/2$ can be identified as the "dis-entanglement" power, which is seen to be zero in the \Moller scattering process.}
While non-stabilizerness is produced in both processes, 
\Moller scattering does not generate  
non-local non-stabilizerness when the initial stabilizer states are entangled, and the outgoing particles remain maximally entangled. This means that the outgoing states are themselves stabilizer states in a different local basis. 
We also observe the same effect in other quantum electrodynamics (QED) processes such as $e^+ e^- \rightarrow \mu^+ \mu^-$, and, it was also recently found in gluon, gluino, graviton and gravitino scattering~\cite{Gargalionis:2026onv}. Since, for 2-qubit systems, maximally entangled states are also stabilizers states, the interaction would need to disentangle the system by some amount to introduce non-local non-stabilizerness, which does not occur in these processes.
More generally, for arbitrary 2-qubit initial states, 
the ability of the interaction to disentangle during those processes is 
very small, when compared to the entangling power. 

In contrast, this feature is not observed in the case of low-energy S-wave 
$np$ scattering, for which the nuclear force
disentangles maximally-entangled particles, and produces non-local non-stabilizerness (or anti-flatness). 
In fact it is shown in Fig.~\ref{fig:np_2x3} of App.~\ref{app:np} that the "dis-entangling power" is about the same or larger than the entangling power in this process, depending on the group of initial states. 
We also find that the non-local non-stabilizerness produced from entangled and unentangled states within group-1 and within 
group-2 is the same, while initial entanglement in states from Group-3 yield greater non-local non-stabilizerness, with a different structure.

To explore this further, we have included the chirality-violating electron mass, and find that the results remain intact.
Hence, this does not result from chirality coinciding with helicity in the massless limit.
The entangled stabilizer states can be written in terms of sums of eigenstates of angular momentum, with good reflection symmetry in the computational basis.   
These sums can be shown to be of states of the same $\hat J_i$ along  one of the Cartesian axes.
The fundamental symmetries in \Moller scattering, 
including identical particles, conservation of total angular momentum and parity,
restrict transformation on the maximally entangled states to tensor-products of single-spin rotations.
These therefore remain maximally entangled.
When considering neutron-neutron scattering (as opposed to $np$ scattering of the previous section), this feature also occurs, due to the identical nature of the particles.

\section{Summary}
\label{sec:summary}
\noindent
Towards a better understanding of the generation of quantum complexity 
in few-body and many-body systems from fundamental interactions,
we have considered the non-local non-stabilizerness and anti-flatness in 
two-particle scattering processes.   
While large-scale non-local non-stabilizerness and anti-flatness in a physical system drive the need for quantum computers in simulations, in two-particle processes they define beyond-classical, basis-independent quantum observables and are fundamental elements of larger systems.  
We have focused on the spin-space structure of two such processes, 
low-energy S-wave 
neutron-proton
scattering and high-energy \Moller scattering.  Starting with each of the tensor-product two-qubit stabilizer states, we have shown that there are a 
small number of  groups of final-states exhibiting the same non-stabilizerness, and whose non-linear non-stabilizerness and anti-flatness are related by a factor of four.
From an experimental standpoint, this relationship is convenient because only the spin of one of the final-state particles is required to be measured from polarized initial states, and not the spin correlations of both particles, as would be required for a CHSH measurement.
We have also found substantial differences in the effectiveness of quantum electrodynamics and nuclear forces to disentangle and induce non-local non-stabilizerness in the scattering processes, when the initial particles are maximally entangled. 

However, this is found to result from the fundamental symmetries of the systems, particularly the additional constraint imposed by the scattering of identical particles, 
as opposed to the structure of the forces themselves.

One may, once again, speculate about the connections between symmetries and the 
minimal entangling power of the S-matrix in confining theories. 
In Ref.~\cite{Beane:2018oxh}, the vanishing entangling power resulting from
identical phase shifts (or phase shifts different by $\pi/2$)
was connected to enhanced emergent spin-flavor symmetries in nucleon-nucleon scattering (Wigner's SU(4) symmetry) and in hyperon-nucleon scattering (SU(16) symmetry).
The latter is a larger group than the SU(6) symmetry group required by the large-N$_c$ limit of QCD, while Wigner's symmetry is the same.
Given the results presented in this work, it is not possible to separate vanishing entanglement power from vanishing non-local non-stabilizing power. 
Given that some of the large-scale entangled states can be prepared efficiently with classical resources, while those with large-scale non-local non-stabilizerness cannot, 
we highlight the fact that minimizing fluctuations in entanglement also minimize fluctuations 
in non-local non-stabilizerness.

\begin{acknowledgements}
\noindent
We would like to thank
Krishna Kumar for enlightening
discussions about MOLLER and other electron scattering experiments.
We are grateful to the organizers and participants of the First~\footnote{\url{https://mbqm.tii.ae/}} and Second~\footnote{\url{https://iqus.uw.edu/events/iqus-workshop-2025-2/}} 
International Workshops on Many-Body Quantum Magic.
This work was supported, in part, by Universit\"at Bielefeld
(Caroline), and
by U.S. Department of Energy, Office of Science, Office of Nuclear Physics, InQubator for Quantum Simulation (IQuS)\footnote{\url{https://iqus.uw.edu}} under Award Number DOE (NP) Award DE-SC0020970 via the program on Quantum Horizons: QIS Research and Innovation for Nuclear Science\footnote{\url{https://science.osti.gov/np/Research/Quantum-Information-Science}} (Martin).
This work was also supported, in part, through the Department of Physics\footnote{\url{https://phys.washington.edu}}
and the College of Arts and Sciences\footnote{\url{https://www.artsci.washington.edu}} at the University of Washington. 
We have made extensive use of Wolfram {\tt Mathematica}~\cite{Mathematica}.
\end{acknowledgements}

\bibliography{biblio_notes}

\clearpage
\onecolumngrid
\appendix
\section{Measurements for Estimating Anti-Flatness}
\label{app:Meas}
\noindent
For a two-particle final state, the reduced density matrix corresponds to a $2\times 2$ matrix that, in spin-space, can be reconstructed from expectation values of the spin operator along (any) three cartesian axes,
\begin{eqnarray}
\rho_A & = & {1\over 2}
\sum_{i=1}^4\  \langle\overline{\sigma}_i\rangle_A\ \ \overline{\sigma}_i
\ \ ,
\label{eq:rebuildrho}
\end{eqnarray}
where 
$\langle\overline{\sigma}_i\rangle_A 
= {\rm Tr} \left(\hat \rho_A \overline{\sigma}_i\right)$.
Measuring the spin components of one of the final state particles, 
provides an estimator for $\rho_A$ from Eq.~(\ref{eq:rebuildrho}),
and hence an estimator for the anti-flatness using Eq.~(\ref{eq:Antiflat}).

\section{Two-qubit Stabilizer States}
\label{app:stabs}
\noindent
For two qubits (with $d=4$), there are four stabilizer operators for each of the 
sixty stabilizer states given in Table~\ref{tab:TwoQstabs}.
Thirty-six of these states are tensor products formed from one-qubit stabilizers, 
while the remaining twenty-four are entangled states.
\begin{table}[!htb]
\centering
\begin{tabularx}{0.5\columnwidth}{c|cccc||c|cccc} 
\hline\hline
state & $|00\rangle$ & $|01\rangle$ & $|10\rangle$ & $|11\rangle$
& state & $|00\rangle$ & $|01\rangle$ & $|10\rangle$ & $|11\rangle$\\
\hline
1 & 1 & 1 & 1 & 1    & 37 & 0 & 1 & 1 & 0 \\
2 & 1 & -1 & 1 & -1  & 38 & 1 & 0 & 0 & -1\\
3 & 1 & 1 & -1 & -1  & 39 & 1 & 0 & 0 & 1 \\
4 & 1 & -1 & -1 & 1  & 40 & 0 & 1 & -1 & 0\\
5 & 1 & 1 & i & i    & 41 & 1 & 0 & 0 & i\\
6 & 1 & -1 & i & -i  & 42 & 0 & 1 & i & 0 \\
7 & 1 & 1 & -i & -i  & 43 & 0 & 1 & -i & 0 \\
8 & 1 & -1 & -i & i  & 44 & 1 & 0 & 0 & -i\\
9 & 1 & 1 & 0 & 0    & 45 & 1 & 1 & 1 & -1\\
10 & 1 & -1 & 0 & 0  & 46 & 1 & 1 & -1 & 1\\
11 & 0 & 0 & 1 & 1   & 47 & 1 & -1 & 1 & 1\\
12 & 0 & 0 & 1 & -1  & 48 & 1 & -1 & -1 & -1\\
13 & 1 & i & 1 & i   & 49 & 1 & i & 1 & -i\\
14 & 1 & -i & 1 & -i & 50 & 1 & i & -1 & i\\
15 & 1 & i & -1 & -i & 51 & 1 & -i & 1 & i\\
16 & 1 & -i & -1 & i & 52 & 1 & -i & -1 & -i \\
17 & 1 & i & i & -1  & 53 & 1 & 1 & i & -i \\
18 & 1 & -i & i & 1  & 54 & 1 & 1 & -i & i \\
19 & 1 & i & -i & 1  & 55 & 1 & -1 & i & i \\
20 & 1 & -i & -i & -1& 56 & 1 & -1 & -i & -i   \\
21 & 1 & i & 0 & 0   & 57 & 1 & i & i & 1 \\
22 & 1 & -i & 0 & 0  & 58 & 1 & i & -i & -1 \\
23 & 0 & 0 & 1 & i   & 59 & 1 & -i & i & -1 \\
24 & 0 & 0 & 1 & -i  & 60 & 1 & -i & -i & 1 \\
25 & 1 & 0 & 1 & 0  \\
26 & 0 & 1 & 0 & 1  \\
27 & 1 & 0 & -1 & 0  \\
28 & 0 & 1 & 0 & -1  \\
29 & 1 & 0 & i & 0  \\
30 & 0 & 1 & 0 & i  \\
31 & 1 & 0 & -i & 0  \\
32 & 0 & 1 & 0 & -i  \\
33 & 1 & 0 & 0 & 0  \\
34 & 0 & 1 & 0 & 0  \\
35 & 0 & 0 & 1 & 0  \\
36 & 0 & 0 & 0 & 1  \\
\hline\hline
\end{tabularx}
\caption{
The complete set of sixty two-qubit stabilizer states.  
For notational purposes, we identify 
$|0\rangle=|\uparrow\rangle$ and $|1\rangle=|\downarrow\rangle$.
The left set are from the tensor product of one-qubit stabilizer states, 
while the right set are entangled states.   
They are (generally) unnormalized, 
and require coefficients of either 1 or ${1\over\sqrt{2}}$ or ${1\over 2}$.
}
\label{tab:TwoQstabs}
\end{table}
%

\section{More Details on Low-Energy NN Scattering}
\label{app:np}
\begin{figure}[!ht]
    \centering
    \includegraphics[width=0.75\textwidth]{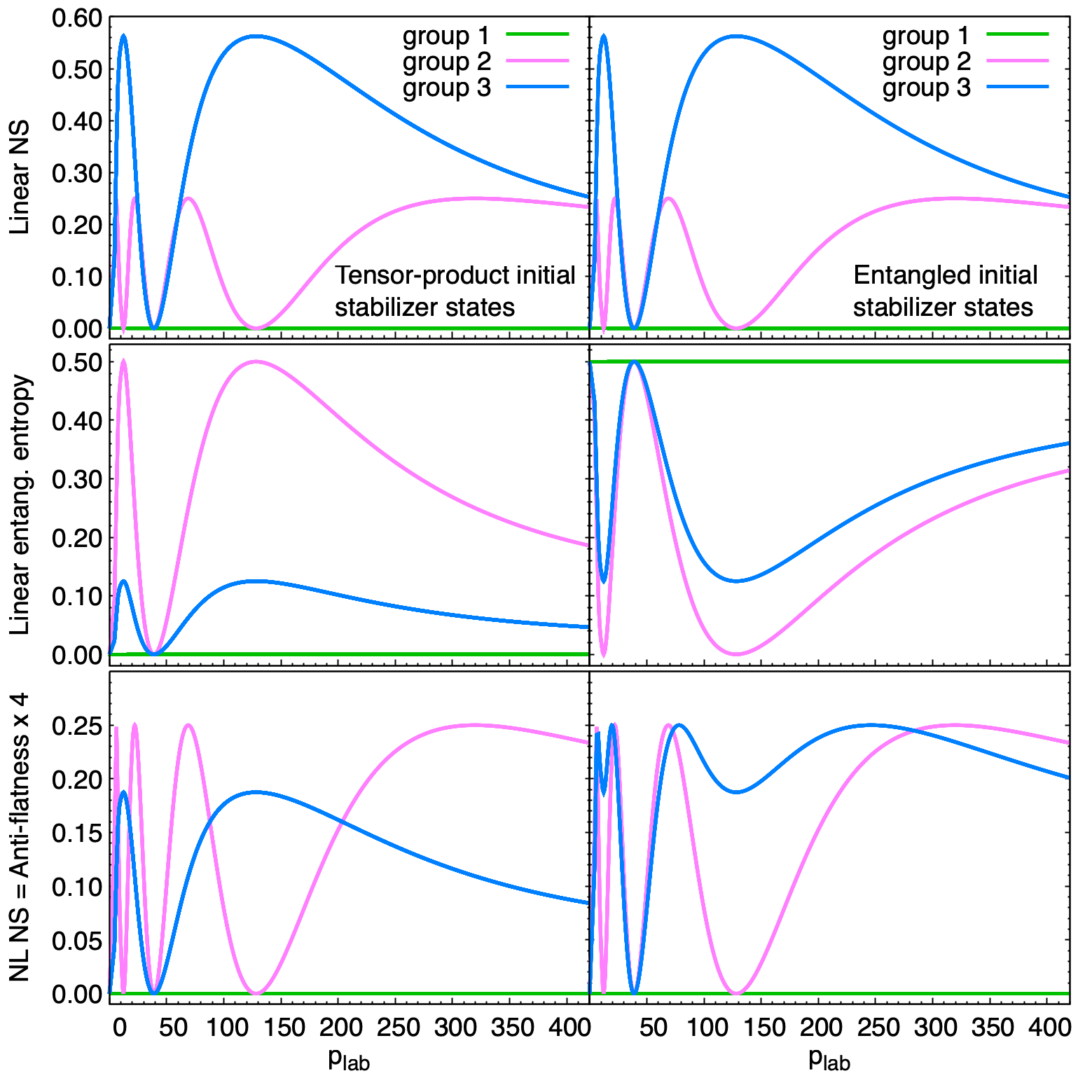}
    \caption{Total linear non-stabilizerness (NS) (top), linear entanglement entropy (middle), and non-local linear non-stabilizerness (NL NS) (anti-flatness $\times 4$) in NN scattering. 
    The left (right) panels show the results for outgoing states from initial unentangled (entangled) stabilizer states.
    }
    \label{fig:np_2x3}
\end{figure}

\noindent
For NN scattering, the stabilizer states separate into three distinct groups, 
where the states in each group lead to the same total linear non-stabilizerness in the outgoing states:
\begin{eqnarray}
&& \text{Group 1} =
\left\{
\begin{array}{l}
1, 4, 17, 20, 33, 36 \text{ (tensor products)} \\
37 \rightarrow 41, 44, 45, 48, 57, 60 \text{ (entangled)} 
\end{array}
\right.\\
&& \text{Group 2} =
\left\{
\begin{array}{l}
2, 3, 18, 19, 34, 35 \text{ (tensor products)} \\
42, 43, 46, 47, 58, 59 \text{ (entangled)}
\end{array}
\right. \\
&& \text{Group 3} =
\left\{
\begin{array}{l}
5 \rightarrow 16 \ {\rm and} \  21 \rightarrow 32 \text{ (tensor products)} \\
49 \rightarrow 56 \text{ (entangled)}
\end{array}
\right.
\label{eq:Groups}
\end{eqnarray}
Fig.~\ref{fig:np_2x3} shows the full linear non-stabilizerness $\mathcal{M}_{\rm lin}(\hat{S} \ket{\psi_i})$, 
linear entanglement entropy and non-local non-stabilizerness (here equivalent to anti-flatness) for outgoing states in NN scattering resulting from tensor-product and entangled initial states $\ket{\psi_i}$ from groups 1, 2 and 3. 

While the total non-stabilizerness produced is the same for tensor-product and entangled initial states within each group, the fluctuations in entanglement differ for group-3. Specifically it is seen that, in this group, the interaction dis-entangles more effectively than it entangles. This results in non-local non-stabilizerness (or anti-flatness) patterns exhibiting more structure when the initial states are entangled. 
This is not the case for states of group-2 for which the dis-entangling power is about the same as the entangling power.

\section{More Details on High-Energy \Moller Scattering}
\label{app:Moll}
\noindent
We choose to use the helicity-amplitude parameterization of the scattering amplitude, with expressions given in Ref.~\cite{10.21468/SciPostPhys.3.5.036,cerveralierta2019thesis},
and not the computational basis used in Ref.~\cite{Liu:2025qfl}.
However, given the basis independence of anti-flatness and non-local non-stabilizerness, either basis could have been chosen.
The helicity amplitudes that are non-zero in the high-energy limit (or $m_e\rightarrow 0$), 
in terms of Mandelstam variables, are~\cite{10.21468/SciPostPhys.3.5.036,cerveralierta2019thesis} 
(neglecting factors of $e^2$, the electromagnetic coupling constant)
\begin{eqnarray}
A_{RR:RR} & = & A_{LL:LL}\ =\ -{2 (t+u)^2\over t u}
\ ,\nonumber\\
A_{RL:RL} & = & A_{LR:LR}\ =\ -{2 u\over t}
\ ,\nonumber\\
A_{RL:LR} & = & A_{LR:RL}\ =\ +{2 t\over u}
\ .
\end{eqnarray}
They can be written in terms of the center-of-momentum scattering kinematics,
\begin{eqnarray}
p_1 & = & (E/2,0,0,|{\bf p}_e|)\ ,\ 
p_2 \ =\  (E/2,0,0,-|{\bf p}_e|)\ \ ,\nonumber\\
p_3 & = &   (E/2,0, |{\bf p}_e| \sin\theta,|{\bf p}_e| \cos\theta)\ ,\ 
p_4 \ =\  (E/2,0,-|{\bf p}_e| \sin\theta,-|{\bf p}_e| \cos\theta)
\ ,\nonumber\\
|{\bf p}_e|^2 & = & (E/2)^2+m_e^2
\ ,\nonumber\\
s & = & (p_1+p_2)^2\ ,\ t\ =\ (p_1-p_3)^2\ ,\ u\ =\ (p_1-p_4)^2
\ ,\nonumber\\
s & = & E^2
\ ,\ t\ =\ -2 |{\bf p}_e|^2 (1-\cos\theta)
\ ,\ u\ =\ -2 |{\bf p}_e|^2 (1+\cos\theta)
\ ,
\end{eqnarray}
where $\theta$ is the scattering angle in the center of momentum.

For \Moller scattering, 
the stabilizer states contributing to the final state non-stabilizerness separate into five distinct groups:
\begin{eqnarray}
&& \text{Group 1} =
\left\{
\begin{array}{l}
33, 36  \text{ (tensor products)} \\
37 \rightarrow 41, 44 \text{ (entangled)} 
\end{array}
\right.\\
&& \text{Group 2} =
\left\{
\begin{array}{l}
1, 4, 17, 20 \text{ (tensor products)} \\
45, 48, 57, 60 \text{ (entangled)}
\end{array}
\right. \\
&& \text{Group 3} =
\left\{
\begin{array}{l}
2, 3, 18, 19 \text{ (tensor products)} \\
46, 47, 58, 59 \text{ (entangled)}
\end{array}
\right. \\
&& \text{Group 4} =
\left\{
\begin{array}{l}
34, 35 \text{ (tensor products)} \\
42 \text{ (entangled)}
\end{array}
\right.\\
&& \text{Group 5} =
\left\{
\begin{array}{l}
5 \rightarrow 16 \ {\rm and} \  21 \rightarrow 32  \text{ (tensor products)} \\
49 \rightarrow 56 \text{ (entangled)}
\end{array}
\right.
\label{eq:GroupsMoll}
\end{eqnarray}
We find that the tensor-product states within each Group-1,2,3,4 display the same anti-flatness. 
However tensor-product states in Group-5 do not.
Hence, we further sub-divide  the tensor-product states in  Group-5 into two sub-groups,
\begin{eqnarray}
    {\rm Group}-5a & = &  5, 6, 7, 8, \ {\rm and} \ 13, 14, 15, 16
    \ ,\nonumber \\
    {\rm Group}-5b & = & 9, 10, 11, 12, \ {\rm and} \ 21 \rightarrow 32
    \ .
\label{eq:GroupsMollGrp3ab}
\end{eqnarray}

\begin{figure}[!ht]
    \centering
    \includegraphics[width=0.75\textwidth]{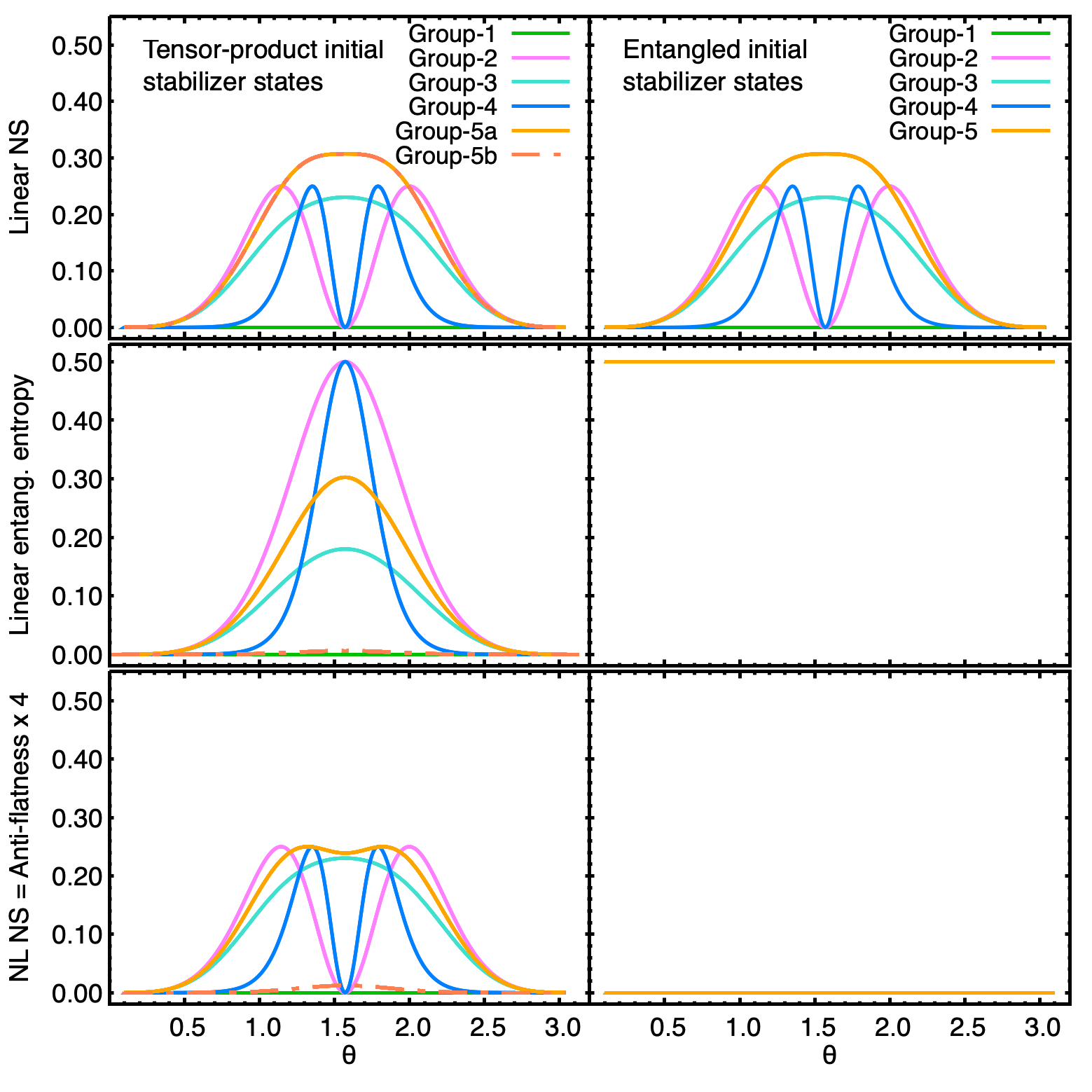}
    \caption{Total linear non-stabilizerness (NS) (top), linear entanglement entropy (middle), and non-local linear non-stabilizerness (NL NS) (equal to anti-flatness $\times 4$) in \Moller scattering. 
    The left (right) panels show the results for outgoing states from initial unentangled (entangled) stabilizer states. }
    \label{fig:Moller_2x3}
\end{figure}

Fig.~\ref{fig:Moller_2x3} shows 
the full linear non-stabilizerness, linear entanglement entropy and non-local non-stabilizerness (here equivalent to anti-flatness) for outgoing states in \Moller scattering $\ket{\chi_i} = \mathcal{N} \hat{\mathcal A} \ket{\psi_i}$ resulting from tensor-product and entangled initial states $\ket{\psi_i}$ from groups 1-5. 
It is interesting to note that the entanglement entropy is maximal at $\theta=\pi/2$ for all groups, and for the average value, 
consistent with the observations in Ref.~\cite{10.21468/SciPostPhys.3.5.036,cerveralierta2019thesis},
while this is not the case for the non-stabilizerness.

In contrast to the strong nuclear force in NN scattering, we observe a large 
asymmetry between the entanglement and dis-entanglement power in the present quantum electrodynamics process. In particular the interaction is not able to decrease the entanglement of maximally-entangled initial states, which results in no non-local non-stabilizerness (or anti-flatness) being produced. Thus in this case the generated non-stabilizerness is entirely local, meaning that the outgoing states remain entangled stabilizer states in a different local basis.

\end{document}